# Novel Advancements in Three-Dimensional Neural Tissue Engineering and Regenerative Medicine

Richard J. McMurtrey[1,2]

[1] Institute of Neural Regeneration & Tissue Engineering, Highland, UT 84003, United States
Email: *richard.mcmurtrey@neuralregeneration.org*

[2] Institute of Biomedical Engineering, Department of Engineering Science, Old Road Campus Research Building, University of Oxford, Oxford, OX3 7DQ, United Kingdom Email: *richard.mcmurtrey@eng.oxon.org*

Keywords: Neural Tissue Engineering, Neural Regeneration, 3D Cell Culture, Organoids, Nanofiber Scaffolds, Hydrogels

Neurological diseases and injuries present some of the greatest challenges in modern medicine, often causing irreversible and lifelong burdens in the people whom they afflict. Conditions of stroke, traumatic brain injury, spinal cord injury, and neurodegenerative diseases have devastating consequences on millions of people each year, and yet there are currently no therapies or interventions that can repair the structure of neural circuits and restore neural tissue function in the brain and spinal cord. Despite the challenges of overcoming these limitations, there are many new approaches under development that hold much promise. Neural tissue engineering aims to restore and influence the function of damaged or diseased neural tissue generally through the use of stem cells and biomaterials. Many types of biomaterials may be implemented in various designs to influence the survival, differentiation, and function of developing stem cells, as well as to guide neurite extension and morphological architecture of cell cultures. Such designs may aim to recapitulate the cellular interactions, extracellular matrix characteristics, biochemical factors, and sequences of events that occur in neurodevelopment, in addition to supporting cell survival, differentiation, and integration into innate neural tissue.

Much work has been done on patterning axonal guidance and neurite outgrowth on two-dimensional (2D) surfaces, but recent work has now demonstrated the ability to pattern neurite extension in three-dimensional (3D) cultures using a combination of functionalized nanofiber scaffolding embedded within hydrogel architecture (McMurtrey, 2014). This study demonstrated that nanofibers could be aligned and patterned using an external scaffolding, and the nanofiber scaffolding could be embedded within a cellularized hydrogel architecture to create composite 3D neural tissue constructs. It was then shown that differentiated neuronal cells within the hydrogel extended neurites that directly tracked along nanofibers, particularly those nanofibers that had been functionalized with laminin and embedded in hyaluronic acid hydrogels, thus resulting in significant overall alignment of neurites with nanofibers in three dimensions. It was also shown that laminin-coated nanofibers in hyaluronic acid hydrogels could significantly enhance the length of neurite extensions. In this study, even non-functionalized polymer nanofibers were able to exert some amount of control over neurite tracking and angle orientation, and although these effects were further enhanced by laminin functionalization of the fibers, these results suggest that topographical cues alone can also influence neurite extension and cell behavior. The mechanisms of this phenomenon are still not well understood, but similar effects have been observed in a variety of cell types (Kim et al., 2012), and in neuronal cells this effect may relate to the ability to extend along radial glial cells in early neurodevelopment. This work suggests that high-resolution patterning of scaffolds within 3D hydrogels is not only possible but provides highly advantageous capabilities in exerting directional control of neurite extension in 3D mediums and producing desired details in the structure of engineered neural tissue.

The function of neural tissue is highly dependent on its intricate 3D architecture at the cellular and subcellular levels, and therefore for neural cells in particular, 3D culture is necessary to form the unique relationship of structure and function in neural tissue. Furthermore, unlike 2D cultures, 3D cultures enable realistic reconstruction of biochemical gradients and interactions among surrounding cells and extracellular matrix. 3D constructs can simulate mass transfer characteristics reminiscent of neural tissue, which may help cells adapt for implantation into the harsh conditions of the body—recent evidence suggests that stem cells implanted into the body may have better survival if they are cultured and prepared under certain conditions of stress or exposure, a form of preconditioning cells (Sart et al., 2014), and culturing cells in 3D conditions with limited diffusion capacities is one way of accomplishing environmental preconditioning. Diffusion characteristics may be modeled for each construct design and may also be improved for diffusion-limited constructs through a variety of means, such as bioreactor design and hydrogel constructs designed with perfusion vents or vascular-like channels.

A particularly useful method for creating 3D cultures involves the use of hydrogels, which provide an aqueous dispersion medium within natural or synthetic matrix molecules. Hydrogels provide the advantage of supporting structural relations among cells, preventing anoikis, and precluding cells from washing away into media or cerebrospinal fluid, while also delivering controllable concentrations of ions, nutrients, and growth factors. Hydrogels generally have a low elastic modulus that can be tuned to simulate the stiffness of neural tissue by means of cross-linking and polymer type and density, which can thereby influence the differentiation of cell lineages (Aurand et al., 2012). These design choices also affect the biocompatibility and degradation rate of the construct, and therefore must be optimized to avoid foreign body reactions and enable biodegradation as constructs are replaced by new tissue formation, but constructs should also endure long enough that they support the survival and integration of newly implanted cells. Hyaluronic acid is one good choice for neural tissue due to its cross-linking capabilities and its natural involvement in neurodevelopment and presence around neurons, and cross-linked hyaluronan hydrogels have already been used to enhance survival of neural stem cells implanted in the central nervous system (Liang et al., 2013; Moshayedi & Carmichael, 2013). Although hyaluronan hydrogels by themselves do not provide high-affinity cell attachment sites, the inclusion of functionalized nanofibers can serve as cell attachment scaffolding that directs neurite outgrowth along the fibers. Conversely, Matrigel possesses abundant cell attachment molecules throughout the hydrogel, which likely explains why lower direct neurite tracking of functionalized fibers seems to occur in Matrigel constructs, and which suggests that hydrogel choice and overall construct design can greatly influence the outcome of neural structure and morphology (McMurtrey, 2014).

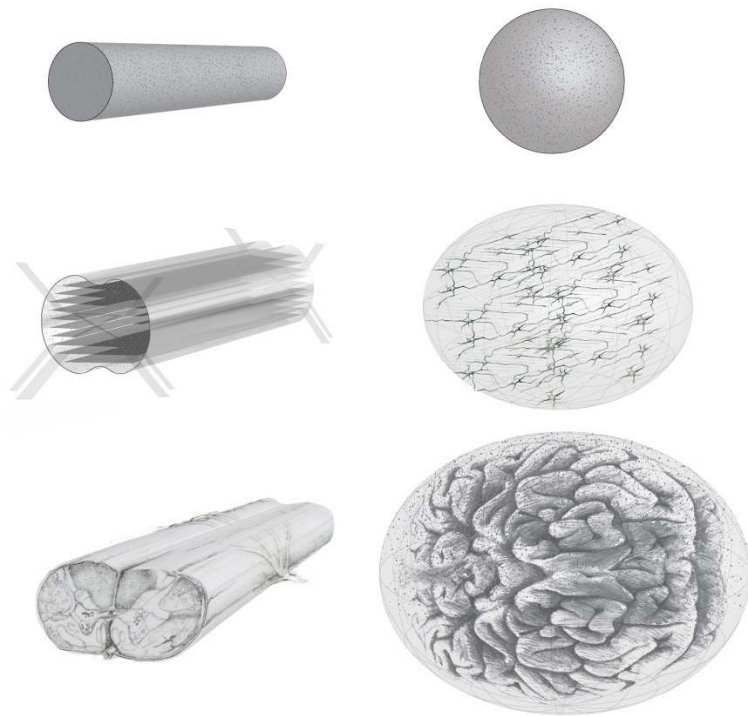

**Figure 1: 3D neural tissue constructs.** Row 1 – Neuronal cultures in 3D hydrogels can be formed in a variety of shapes and designs. Row 2 – Regional identities and neurite extensions can be patterned with signaling factors and nanofiber architecture embedded within the hydrogel. Row 3 – 3D tissue and organoid models will provide incredible new tools and insights into neurodevelopment and neurological injury and disease, as well as great potential for regenerating functional neural tissue from stem cells. [Images used with permission of the Institute of Neural Regeneration & Tissue Engineering.]

Early attempts at creating neural tissue from pluripotent stem cells focused primarily on the optimal combination of signaling factors in 2D cultures, but many of these attempts were only partially successful, with neural tissue failing to achieve proper cellular structure, organize synaptic networks, express complete molecular markers, or mature in electrophysiological function (Hansen et al., 2011). The maturation from human pluripotent stem cells into cell types from all layers of the adult human cortex was then demonstrated in 2D cultures, wherein the maturation process was shown to recapitulate that which occurs in normal development, with deep layer VI neurons maturing first and upper layer II/III neurons later, then finally the differentiation of astrocytes and formation of functional synaptic connections reminiscent of early innate cortical networks (Shi et al., 2012). Subsequently, 3D hydrogel cultures of pluripotent stem cells were shown to enable the formation of neuroanatomical structures that are simply not possible in 2D cultures, and it seems that the cells themselves possess innately-programmed capabilities to self-assemble at least some aspects of neuroanatomical structures even in unpatterned hydrogel constructs (Lancaster et al., 2013). These structures comprised ventricular, hippocampal, retinal, and cortical regions, including spatially separated upper and lower layers of cortex, and, owing to the observed multicellular differentiation and spatial organization characteristic of whole organs, these structures have been termed "cerebral organoids." Although these constructs can serve as models of organogenesis, fully mature neural structures along with the full composition of central nervous system cell types still remain incomplete or only partially developed, and although the neurons in these 3D organoids were shown to develop spontaneous activity, further synaptic characterization and network functionality has yet to be examined. The formation of functional networks is dependent on excitatory neurons, inhibitory neurons, and glial cell types, and because these cell types arise from different regions or at different stages of development, enabling the desired cell types to form appropriate spatial structures at appropriate times is a difficult challenge. It is not yet known to what extent neural networks can be functionally established in 3D constructs, or whether 3D constructs can be used to help integrate functional neural networks into the prohibitive *in vivo* central nervous system environment, but the ability to replicate neuroanatomical structures and guide neural network formation in three dimensions, both *in vitro* and *in vivo*, would be a major advance in the field of neural regeneration.

The formation of neuroanatomical structures at the cellular and subcellular levels will likely depend on a combination of cellular, biochemical, and structural cues that can guide desired axis patterning, regional identity, and cellular architecture. Because hydrogels and nanofibers can provide these components, an ideal approach for reconstructing structural and functional components of neural tissue may therefore include the integration of scaffolding and hydrogel into singular 3D constructs. Specific gradients and concentrations of signaling factors can be integrated into hydrogels, and nanofibers can be coated with a variety of factors that influence cell attachment, survival, growth, differentiation, migration, and neurite extension. Nanofibers may also be functionalized with pharmacologic agents that elute and diffuse from the fibers to create localized gradients within the hydrogel, or agents that are chemically coupled to the fibers so that regional concentrations remain tightly controlled. Such composite hydrogel constructs may be functionalized to minimize foreign-body reactions and enhance integration into neural tissue, and the hydrogel architecture also provides stabilizing protection of neuroglial cell cultures during early structural development and during stages of transplantation and grafting. Also, it has been recognized that adverse side effects of pain or spasticity might result from some spinal cord regeneration effects (Tuszynski & Steward, 2012), but targeted connections of axonal and dendritic pathways and suppression of unwanted connections might be achieved through the use of patterned scaffolding that can guide neural pathways at the cellular and subcellular levels. Even unpatterned fibrin scaffolds have shown significant capabilities in enhancing functional axonal extensions across long-distance lesions in rat spinal cord injuries (Lu et al., 2012), which further suggests that the integration of cellularized hydrogels and patterned scaffolding constructs may provide significant benefits for successful reconstruction of functional neural tissue in patients.

The ability to use a patient's own pluripotent stem cells to replicate regional components of neurodevelopment and rebuild neuroanatomical structures in 3D will not only provide a tremendous new tool for the successful implementation of regenerative medicine, but will also enable detailed study of complex neurodevelopmental processes, neurite guidance systems, and pathological conditions, including neurodegenerative diseases (Choi et al., 2014) and neurogenetic diseases (Lancaster et al., 2013), particularly for diseases that are not easily characterized in animals and for processes and neuroanatomical organizations that appear unique in humans, such as regions of neural progenitor zones and activity of radial glia. These 3D constructs can also serve as useful models of traumatic injury and pharmacological or toxicological testing. The concept of using functionalized scaffolding within hydrogels for neural tissue architecture can now be scaled up to create more intricate structures that better replicate the complexity of innate neuroanatomical structures, such as those of the cortex, spinal cord, hippocampus, or other regions. For example, the formation of multi-layered cortex may be guided with layered nanofibers, with upper layers designed to promote intracortical connections and lower layers functionalized and shaped to promote deep projections. Likewise, hippocampal anatomy may be formed with hydrogel architecture and nanofiber scaffolding that guides perforant, collateral, and commissural pathways. Similarly, spinal cord anatomy could be reconstructed with specific tracts and decussations guided by nanofiber scaffolding through hydrogel. With the use of induced pluripotent stem cells or induced neuronal cells, these 3D constructs could then be studied as regional models of the central nervous system or could one day be implemented as autologous grafts into damaged sites of the nervous system in order to restore neural function, particularly for damaged sites of spinal cord, areas of stroke infarction, tumor resection sites, peripheral nerve injuries, or areas of neurodegeneration.